\documentclass[twocolumn,prb,amsmath,superscriptaddress]{revtex4}
\usepackage{graphicx}
\newcommand{\be}{\begin{equation}}
\newcommand{\ee}{\end{equation}}
\newcommand{\ber}{\begin{subequations}\begin{eqnarray}}
\newcommand{\eer}{\end{eqnarray}\end{subequations}}
\newcommand{\berb}{\begin{eqnarray*}}
\newcommand{\eerb}{\end{eqnarray*}}

\newcommand{\mH}{{\mathcal H}}
\newcommand{\mI}{{\mathcal I}}

\newcommand{\mC}{{\mathcal C}}

\newcommand{\ket}[1]{\left|#1\right\rangle}
\newcommand{\bra}[1]{\left\langle#1\right|}
\newcommand{\tr}{{\rm Tr}}
\begin{document}
\title{Extremal asymmetric universal cloning machines}
\author{Mingming Jiang} 
\affiliation{Department of Modern Physics, University of Science
and Technology of China, Hefei 230027, People's Republic of China}
\affiliation{Laboratory of Excited State Process, Changchun Institute of Optics, Fine Mechanics, and Physics, Chines Academy of Science, Changchun 130033, People's Republic of China}
\author{Sixia Yu}
\affiliation{Department of Modern Physics, University of Science
and Technology of China, Hefei 230027, People's Republic of China}
\begin{abstract}
The trade-offs among various output fidelities of asymmetric
universal cloning machines are investigated. First we find out all the attainable
optimal output fidelities for the  1
to 3 asymmetric universal cloning machine and it turns out that
there are two kinds of extremal asymmetric cloning machines which have to
cooperate in order to
achieve some of the optimal output fidelities. Second we construct a family 
of extremal cloning machines that includes the universal symmetric cloning machine
as well as an asymmetric 1
to $1+N$ cloning machine for qudits with two different output
fidelities such that the optimal trade-off between the measurement
disturbance and state estimation is attained in the limit of
infinite $N$. %\pacs{03.67.-a, 03.65.-w, 42.50.Dv}
\end{abstract}
\maketitle

A single quantum can neither be cloned \cite{nature} nor be
broadcasted \cite{broad}, but it can be approximately cloned
universally for qubits \cite{qubit1,qubit2} and  for qudits
\cite{qudit1, werner, fan}, or probabilistically \cite{duan},
symmetrically or asymmetrically \cite{niu,cerf1,cerf2}, and
experimentally \cite{exp}. The quantitative boundary between what is
possible and impossible hinted by the no-cloning theorem is rarely
explored apart from a few cases including the optimal symmetric
cloning machines \cite{niu,werner}, and the optimal $1 \mapsto 2$
and $1 \mapsto 3$  asymmetric cloning machines \cite{m}.

A universal $1\mapsto N$ cloning machine is a quantum
mechanical process with one input and $N$ outputs with the fidelity
between each output state and the input state being independent of
the input state. Symmetric cloning machines, which are special cases
of asymmetric cloning machines, are characterized by the unique
maximal attainable output fidelity. For asymmetric cloning machines
optimal trade-offs among the output fidelities in certain range of
values have been explored \cite{m}. In addition, a $1$ to $1+n$ asymmetric cloning machine
with 2 different output fidelities for qubits  has also been
constructed which,  in the large $n$ limit, balances the inequality
of measurement disturbance and state estimation \cite{ineq}.

In this letter we shall present at first the complete trade-off of
output fidelities of  $1$ to $3$ cloning machine for qudits. It turns out that
there are two kinds of extremal cloning machines and for some range of output fidelities the two extremal cloning
machines must cooperate to attain the optimal fidelities instead of
a single ``optimal" cloning machine. Second we construct also a $1$
to $1+n$ cloning machine for qudits, which belongs to a family of extremal cloning machines in the symmetric subspace, that saturates Banaszek's
inequality of measurement disturbance and state estimation. 

In the following we consider only qudits, i.e., $d$-level systems
whose Hilbert space is spanned by $\{\ket n\}_{n=0}^{d-1}$. Let us
start with a trivial case to establish some notations, namely a 1 to
1 universal cloning  machine, which can be represented by a
completely positive map $\psi\mapsto\mC_1(\psi)$, where $\psi$
represents the density matrix of a pure state $\ket\psi$ of a single
qudit which is labeled by $A$. The output fidelity, taking into
account of the universality, reads
\begin{equation}
F_A=\int \tr(\psi\mC_1(\psi))d\psi
=\frac{d+f_A}{d(d+1)},
\end{equation}
where $f_A=\tr(Q_{RA}\Phi_{RA})$ with
$Q_{RA}=\mI_R\otimes\mC_1(\Phi_{RA})$ being a subnomalized state
($\tr Q_{RA}=d$) of the composite system of a reference qudit $R$
and the original qudit $A$ and  $\Phi_{RA}$ denoting the density
matrix of a (subnormalized) maximally entangled state
$\ket\Phi=\sum_n\ket{nn}$ of the composite system $RA$. It is
obvious that the output fidelity $F_A$ ranges from $1/(d+1)$ to $1$
because $f_A$ takes values from 0 to $d^2$. The maximal output
fidelity arises from the identity map $\mI(\psi)=\psi$ and the
minimal fidelity arises from the fact that the cloning machine must
be a physical process allowed by the principle of quantum mechanics,
i.e., $\mC(\psi)$ is a completely positive map. In the case of $d=2$
the minimal output fidelity is achieved by the optimal universal NOT
gate.

The situation is similar for cloning machines producing two or more
copies. Let us consider now a $1 \mapsto 2$ universal cloning
machine, which can be represented by a completely positive map
$\mC_{2}$ from $\mH_A$ to $\mH_A\otimes \mH_B$. Its two output
fidelities $F_A$ and $F_B$ are determined by the expectation values
$f_A$ and $f_B$ of two observables ${\Phi_{RA}}$ and ${\Phi_{RB}}$
in the subnormalized state $Q_{RAB}=\mI_R\otimes\mC_{2}(\Phi_{RA})$.
Thus the bound of the optimal output fidelities is bounded by all possible expectation
values of two observables ${\Phi_{RA}}$ and ${\Phi_{RB}}$ when the state runs over all possible
states of composite system $RAB$.

Obviously the range of two observables ${\Phi_{RA}}$ and ${\Phi_{RB}}$ is spanned by $2d$ states $\ket\Phi_{RA}\ket k_B$ and
$\ket\Phi_{RB}\ket k_A$ with $k=0,1,\ldots,d-1$, from which an orthonormal basis can be constructed
\begin{equation}
\ket{\phi^\pm_k}=\frac1{\sqrt{2(d\pm1)}}\Big(\ket\Phi_{RA}\ket k_B\pm\ket\Phi_{RB}\ket k_A\Big).
\end{equation}
It is not complete thus $\sum_k{\bf \phi}^+_k+\phi^-_k\le {\bf I}_3$ where $\phi^\pm_k$ denotes
the projector of the corresponding state and ${\bf I}_3$ is the identity matrix for 3-qudit.
When averaged in an arbitrary 3-qudit state $Q_{RAB}$ with normalization $\tr Q_{RAB}=d$
the incompleteness condition leads to
\begin{equation}\label{c2}
\frac{(\sqrt{f_A}+\sqrt{f_B})^2}{2(d+1)}+\frac{(\sqrt{f_A}-\sqrt{f_B})^2}{2(d-1)}\le d.
\end{equation}
This (well-known) inequality can be regarded as an uncertainty relationship
between observables $\Phi_{RA}$ and $\Phi_{RB}$. The expectation values
that saturate the inequality Eq.(\ref{c2}) for a 3-qudit state correspond to the optimal 1 to 2 asymmetric
cloning machine without the restriction that the coefficients be non-negative. Thus the trade-off between
two output fidelities  $F_A$ and $F_B$ can be plotted as
in Fig.1. It should be pointed out that given one of the output
fidelities in the interval between $1-1/d(d+1)$ and $1$ the other
output fidelity assumes a minimal value which is greater than the
minimal possible fidelity $1/(d+1)$.
\begin{figure}
\includegraphics{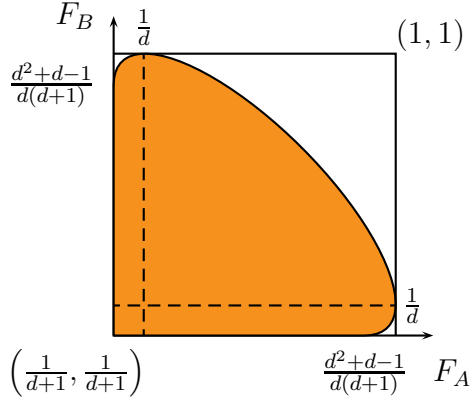}
\caption{(Color online) The trade-off between two output fidelities of 1 to 2 asymmetric cloning
machine. The shaded area which is bounded by two axes and part of
a ellipse contains all possible output fidelities.}
\end{figure}

Let us now consider a 1 to 3 asymmetric universal cloning machine,
which can be represented by a quantum operation $\mC_3$ with 1 input
and 3 outputs. In this case three output fidelities $F_A,F_B$, and
$F_C$ are determined though Eq.(1) by the expectation values
$f_A,f_B$, and $f_C$ of three observables $\Phi_{RA}$, $\Phi_{RB}$,
and $\Phi_{RC}$  in a 4-qudit state
$Q_{RABC}=\mI_R\otimes\mC_3(\Phi_{RA})$ which is subnormalized as
$\tr Q_{RABC}=d$. To explore all the possible output fidelities, we shall at first
find out all the possible expectation values of those three observables in the same state and then
we construct symmetric cloning machines that attain those optimal values.

At first we notice that the Hilbert space of
4-qudit can be decomposed into three orthogonal subspaces
\begin{equation}
{\mH}_4=V_+\oplus V_-\oplus V_0,
\end{equation}
where the supbspace $V_0$ is the orhtogonal complement of $V_+\oplus V_-$ with
subspaces $V_\pm$ spanned by, respectively, by bases $(a=0,1,2)$
\begin{equation}
\ket{\phi_{kl\pm}^{a}}=\frac{{\bf I}+\omega^a {\bf
Y}+\omega^{2a}{\bf  Y}^2} {\sqrt{3(d\pm(3\delta_{a0}-1))}}
\ket\Phi_{RA}\ket{\{kl\}_\pm}_{BC},
\end{equation}
where ${\bf I}_4$ is the identity operator for 4-qudit
and  ${\bf Y}$ denotes  the cyclic permutation operator acting only
on three qudits $A,B,C$ with effects ${\bf
Y}\ket{m,n,k}_{ABC}=\ket{k,m,n}_{ABC}$ for arbitrary $m,n,k$ and
leaving the qudit $R$ unchanged, and
$\ket{\{kl\}_\pm}=(\ket{kl}\pm\ket{lk})/\sqrt2$ for $k>l$ and
$\ket{\{kk\}_+}=\ket{kk}$. 

Subspace $V_+\oplus V_-$ is the range of
three observables $\Phi_{RA}$, $\Phi_{RB}$, and $\Phi_{RC}$ and therefore all the expectation values of these three observables are zero in $V_0$. Furthermore, we have 
$\bra{\phi_{kl+}^{a}}\Phi_{R\alpha}\ket{\phi_{mn-}^{b}}=0$
$(\alpha=A,B,C)$. As a result all the attainable expectation values
of three observables $\Phi_{R\alpha}$ $(\alpha=A,B,C)$ are those convex combinations
of these attainable values in pure states in $V_\pm$ and 0, the value attained in $V_0$. In other words if we
have found out two sets of all the attainable expectation values under the pure states in subspaces $V_\pm$
then the complete set of attainable values is the convex hull of these two sets and 0.

For an arbitrary pure (subnormalized) state $|\psi_\pm\rangle$ in $V_\pm$ with $\langle\psi_\pm|\psi_\pm\rangle=d$ we denote $f_{\alpha\pm}=\langle\psi_\pm|\Phi_{R\alpha}|\psi_\pm\rangle$ for $\alpha=A,B,C$ and $\bf f_{A\pm}$ as a $d(d\pm1)/2$-dimensional complex vector whose components are $\langle\psi_\pm|\Phi\rangle_{RA}|{kl}_\pm\rangle_{BC}$ with $k,l=0,1,\ldots,d-1$ and similarly for $\bf f_{B\pm}$ and $\bf f_{C\pm}$. Obviously $f_{\alpha\pm}=|\bf f_{\alpha\pm}|^2$ for all $\alpha=A,B,C$. Since $V_+\oplus V_-$ is only a subspace one has
\begin{equation}
\sum_{a=0}^2\sum_{k\ge l}^{d-1}\Big(|\phi^{a}_{kl+}\rangle\langle\phi^{a}_{kl+}|+|\phi^{a}_{kl-}\rangle\langle\phi^{a}_{kl-}|\Big)\le {\bf I}_4
\end{equation}
which leads to
\begin{equation}\label{f3}
f_{A\pm}+f_{B\pm}+f_{C\pm}\mp\frac{|{\bf f}_{A\pm}+{\bf f}_{B\pm}+{\bf f}_{C\pm}|^2}{d\pm2}\le d(d\mp1)
\end{equation}
when averaged in the state $|\psi_\pm\rangle$, respectively. Given the lengths of
three complex vectors ${\bf f}_A$, ${\bf f}_B$, and ${\bf f}_C$, the length
$|{\bf f}_A+{\bf f}_B+{\bf f}_C|$ is bounded above by  $|{\bf f}_A|+|{\bf f}_B|+|{\bf f}_C|$
and bounded from below by the maximum among 0, $|{\bf f}_A|-|{\bf f}_B|-|{\bf f}_C|$,
$|{\bf f}_B|-|{\bf f}_A|-|{\bf f}_C|$, and $|{\bf f}_C|-|{\bf f}_B|-|{\bf f}_A|.$
Thus it follows from Eq.(\ref{f3}) that
\be\label{e0}
x^2+y^2+z^2-\frac{(x+y+z)^2}{d+2}\le d(d-1)
\ee
in the symmetric subspace $V_+$, where we have denoted $x=\sqrt{f_A}$, $y=\sqrt{f_B}$, and $z=\sqrt{f_C}$ for convenience, and in the antisymmetric subspace $V_-$ the expectation values satisfy either any one of the following inequalities
\ber
x^2+y^2+z^2+\frac{(x+y-z)^2}{d-2}\le d(d+1)\label{e1}\\
x^2+y^2+z^2+\frac{(x-y+z)^2}{d-2}\le d(d+1)\label{e2}\\
x^2+y^2+z^2+\frac{(x-y-z)^2}{d-2}\le d(d+1)\label{e3}
\eer
together with restrictions $z\ge x+y$, $y\ge x+z$, and $x\ge z+y$, respectively, or lie within the sphere
\be\label{qiu}
x^2+y^2+z^2\le d(d+1)
\ee
restricted by the conditions
\begin{equation}
x\le y+z,\ y\le x+z,\ z\le x+y.
\end{equation}
These bounds specify the range of all the possible expectation values of $\Phi_{R\alpha}$ $(\alpha=A,B,C)$ in pure states. Thus all the possible expectation values of three observables $\Phi_{R\alpha}$ $(\alpha=A,B,C)$ in arbitrary states are all the possible convex combinations of those bounds, i.e., the boundary is the convex hull of those four ellipsoids defined in Eq.(\ref{e0}) and Eqs.(\ref{e1})-(\ref{e3}) and the partial sphere in Eq.(\ref{qiu}), which is explicitly plotted in the Fig.2. We note that the restricted sphere Eq.(\ref{qiu}) is contained in the convex hull for $d\ge 3$ and in the case of $d=2$ the boundary is the convex hull of Eqs.(\ref{e0}) and (\ref{qiu}). Since the function $\sqrt x$ is a one-to-one concave function, the boundary for the fidelities $F_\alpha$ has essentially the same structure as the boundary for $\sqrt {f_\alpha}$ $(\alpha=A,B,C)$.
\begin{figure}
\includegraphics[scale=0.95]{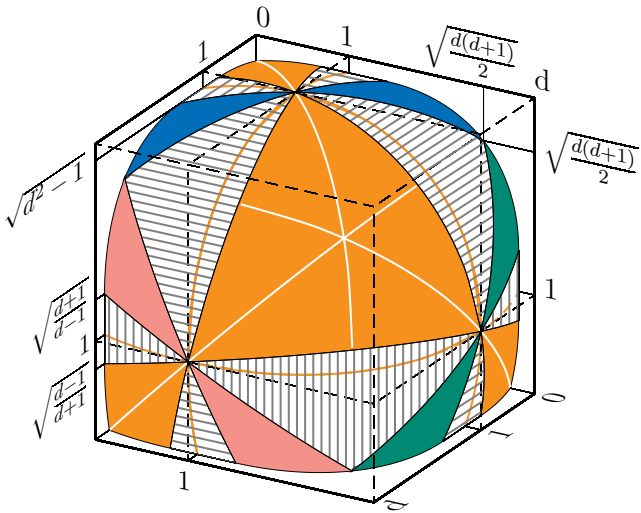}
\caption{(Color online) The convex hull of 4 ellipsoids with colored parts being the extremal points. Three axes are $x=\sqrt{f_A}$, $y=\sqrt{f_B}$ and
$z=\sqrt{f_C}$.}
\end{figure}

In the following we shall prove that the surface of the convex hull as plotted in Fig.2 is attainable by explicitly constructing the universal cloning machines with the desired output fidelities. To do so we have only to construct the cloning machines that saturate those four inequalities Eqs.(\ref{e0}) and (\ref{e1})-(\ref{e3}), respectively. We consider a system of five qudits labeled with $A,B,C,E$, and $F$ and define two unitary evolutions as
\begin{eqnarray}
U_\pm|m_A0_{BCEF}\rangle=\sqrt{\frac{2}{d(d\pm1)}}\Big(\alpha+\beta{\bf Y}+\gamma{\bf Y^2}\Big)\cr|m\rangle_A\Big(\ket\Phi_{BE}\ket\Phi_{CF}\pm\ket\Phi_{CE}\ket\Phi_{BF}\Big),
\end{eqnarray}
where $\bf Y$ is the cyclic permutation acting on $ABC$ as before and $\alpha,\beta,$ and $\gamma$ are real numbers satisfying
\be
\alpha^2+\beta^2+\gamma^2\pm\frac2d(\alpha\beta+\beta\gamma+\gamma\alpha)=1.
\ee
It is easy to check that the cloning machines defined by $U_\pm$ are universal. For convenience we denote $x_\pm=d\alpha\pm(\beta+\gamma), y_\pm=d\beta\pm (\alpha+\gamma)$, and $z_\pm=d\gamma\pm (\beta+\alpha)$.

We consider at first the cloning machine $U_+$. In the case of $x_+,y_+,z_+\ge0$ we have $f_A=x_+^2$, $f_B=y_+^2$, and $f_C=z_+^2$ and the inequality (\ref{e0}) becomes an equality. Thus we have constructed an {\it extremal} cloning machine $U_+$ that saturates the inequality (\ref{e0}). As will see in the following discussions the extremal cloning machines do not always produce the optimal output fidelities. In the case of non-negative $\alpha,\beta,$ and $\gamma$ the unitary evolution $U_+$ defines exactly the asymmetric cloning machine investigated in Ref.\cite{m} with optimal output fidelities corresponding to the central golden area in Fig.2. In the case of two negative and one positive coefficients among $\alpha,\beta,$ and $\gamma$ while keeping $x_+$, $y_+$, $z_+$ non-negative, $U_+$ also gives rise to the optimal cloning machines with fidelities corresponding to three small golden areas in Fig.2. The boundaries of those four golden regions are the intersections between the golden ellipsoid defined by Eq.(\ref{e0}) with planes $(d+1)x=y+z$, $(d+1)y=y+z$, and $(d+1)z=(x+y)$.

Next we consider the cloning machine $U_-$. Three output fidelities of the cloning machine $U_-$ are $f_A=x_-^2$, $f_B=y_-^2$, and $f_C=z_-^2$, and they saturate the inequality Eq.(\ref{e1}) in the case of $x_-,y_-\ge 0$, and $z_-\le 0$. Similarly the inequalities Eqs.(\ref{e2}) and (\ref{e3}) are saturated by choosing $x_-,z_-\ge 0$, and $y_-\le 0$ or $y_-,z_-\le 0$ and $x_-\le 0$. These cloning machines therefore attain the optimal fidelities in the blue, green, and red regions in Fig.2.

In the stripped white regions in Fig.2 the optimal output fidelities are attained by neither of these two extremal cloning machines $U_\pm$. Instead the optimal values can be achieved by a suitable cooperation of $U_\pm$. Since any value in the stripped white regions is a convex combination of the extremal values in the colored regions, it can be attained by mixing properly those extremal cloning machines achieving the extremal values. For example, let $(x,y,z)=p(x,y,z)_G+(1-p)(x,y^\prime,z)_B$ be an optimal value in a stripped white region, that is a convex combination of two optimal values in the blue and golden regions. Let $U_G$ and $U_B$ be the extremal machines described above then by applying the machine $U_G$ with probability $q$ and $U_B$ with probability $1-q$ we obtain the desired optimal fidelity $(x,y,z)$ where $q$ is uniquely determined by $(qy+(1-q)y^\prime)^2=py^2+(1-p)y^{\prime2}$.

At last we consider 1 to $N$ asymmetric universal cloning machines which can be represented
by a quantum operation $\mC_N$ with one input and $N$ outputs which are labelled from
1 to $N$. Each output fidelity
$F_n$ is determined though Eq.(1) by the expectation value $f_n$  of observable
$\Phi_{0n}$ in the subnormalized state $Q_{0N}=\mI_0\otimes\mC_N(\Phi_{01})$. (The reference qubit is labeled with 0.) 
In what follows we shall find out a partial bound for the expectation values of $\Phi_{0k}$ (and therefore output fidelities) and construct the cloning machine attaining this bound. A complete bound even in the simplest case $N=4$ is unattainable so far.

The
range of $N$ observables $\Phi_{0k}$ is spanned by the following $Nd^{N-1}$ states (not normalized):
\be
|\psi^a_\lambda\rangle=\mathbf P_a\ket\Phi_{01}\ket{\lambda}_{23\ldots N},\quad \mathbf P_a=\frac1{N}\sum_{k=0}^{N-1}\omega^{ka}{\bf X}^k
\ee
where ${\bf X}$ is the cyclic permutation acting on $N$ qudits according to
${\bf X}\ket{n_1,n_2,\ldots,n_N}=\ket{n_2,n_3,\ldots,n_N,n_1}$, $a=0,1,\ldots,N-1$ and $\{|\lambda\rangle\}$
is an arbitrary basis for $N-1$ qudits.
Let $\ket\psi$ be an arbitrary pure $(N+1)$-qudit state the Gramm matrix
of these $Nd^{N-1}+1$ states $\{|\psi\rangle,\ket{\psi^a_\lambda}\}$ is semi-positive definite, i.e.,
\begin{equation}
\left(\begin{matrix}
d                &{\bf f}_1         &{\bf f}_2     &{\bf f}_3     &\cdots&{\bf f}_N\\
{\bf f}_1^\dagger&\tr_1{\bf P_0}    & 0 &0&\cdots&0\\
{\bf f}_2^\dagger&0&\tr_1{\bf P}_{1}&0  &\cdots&0\\
{\bf f}_3^\dagger&0&0&\tr_1{\bf P}_2&\cdots&0\\
\vdots&\vdots&\vdots&\vdots&\ddots&\vdots\\
{\bf f}_N^\dagger&0&0&0&\cdots&\tr_1{\bf P}_{N-1}
\end{matrix}\right)\ge0,
\end{equation}
where $\mathbf f_{a+1}$ denotes a $d^{N-1}$-dimensional vector with components $\langle\psi\ket{\psi^a_\lambda}$ for $a=0,1,\ldots,N-1$.

By partitioning the Hilbert space of the last $N-1$ qudits into symmetric subspace, which is spanned by all the symmetric states $|\bf n\rangle_{23\ldots N}$, and its orthogonal complement, the Gramm matrix assumes a quasidiagonal form, and in the symmetric subsapce the non-negativeness of the Gramm matrix gives rise to
\be\label{de0}
\sum_{k=1}^N {f}_k-\frac1{d+N-1}{\left(\sum_{k=1}^N\sqrt{ f_k}\right)^2}\le d(d-1)
\ee
by noticing $N\tr_1{\bf P}_0=d+N-1$ while $N\tr_a{\bf P}_0=d-1$ $(a\ne 1)$ in the symmetric subspace. Here we have denoted $f_k=\langle\psi|{\Phi}_{0k}|\psi\rangle$.

Let us now construct the cloning machine that saturates
the inequality above. Consider the unitary evolution defined by
\begin{eqnarray}
U_\alpha|m\rangle_1|0\rangle_{23\ldots N}|0\rangle_{2^\prime3^\prime\ldots N^\prime}
=\cr\sum_{a=0}^{N-1}\frac{\alpha_a{\bf X}^a}{\sqrt{\binom{d+N-1}N}}\ket m_{1}\sum_{\mathbf n}\ket{\bf n}_{23\ldots N}\ket{\bf n}_{2^\prime3^\prime\ldots N^\prime}
\end{eqnarray}
with real numbers $\alpha_a$ satisfying
\begin{equation}
\sum_{a=0}^{N-1}\alpha_a^2+\frac2d\sum_{a>b}^{N-1}\alpha_a\alpha_b=1.
\end{equation}
As long as $x_{a+1}=(d-1)\alpha_a+\sum_a\alpha_a\ge 0$ for all $a=0,1,\ldots,N-1$, the inequality Eq.(\ref{de0}) is saturated with fidelities given by $f_a=x_a^2$. Obviously the symmetric universal 1 to $N$ cloning machine is a special case.

In addition if we take $\alpha_a=\beta/(d+N-1)$ for $a=1,2,\ldots,N-1$ and $\alpha_0=\alpha+\beta/(d+N-1)$ with $\alpha,\beta$ being non-negative, there are only two different output fidelities $f=(d\alpha+\beta)^2$ and $g=(\alpha+\beta)^2$. The normalized condition, Eq.(18), yields
\begin{equation}
(\sqrt f-\sqrt g)^2=(d-g)(d-1)+\frac{(d\sqrt g-\sqrt f)^2}{d+N-1},
\end{equation}
which saturates the optimal trade-off between the information gain and state disturbance \cite{ineq} when $N$ tends to infinity. The last $N-1$ outputs with the same fidelity $g$ provide the information gain because of the equivalecy between the state estimation and symmetric cloning with an infinite outputs \cite{bruss}, while the first output fidelity $f$ characterizes the disturbance suffered in estimating the quantum state.

It should be pointed out that Eq.(16) needs not to be satisfied by all the optimal output fidelities. That is to say, there are some output fidelities that will fall outside the hype-ellipsoild given by Eq.(16). Therefore, the cloining machine $U_\alpha$ does not always produce the optimal output fidelities. We believe that (without proof) when $\alpha_a\ge 0$ $(a=0,1,\ldots,N-1)$ the asymmetric cloning machine $U_\alpha$ is optimal which means Eq.(16) holds ture for this special range of output fidelities.

We acknowledge the financial support of NNSF of China (Grant No. 10675107).

\end{document}